\documentclass[final,leqno,onefignum,onetabnum]{siamltex1213}
\usepackage[T1]{fontenc}
\usepackage[latin9]{inputenc}
\usepackage[letterpaper]{geometry}
\usepackage{amstext}


\usepackage{graphicx}
\usepackage{caption}
\usepackage[fleqn]{amsmath}
\usepackage{amssymb}

\usepackage{listings}
\usepackage{url}

\def\fatx{\mathbf{x}}
\def\fatmu{\boldsymbol{\mu}}
\def\fatnu{\mathbf{\nu}}

\def\packagename {MOLNs}

\begin{document}

\title{MOLNs: A cloud platform for interactive, reproducible and scalable spatial stochastic computational experiments in systems biology using PyURDME\thanks{This work was funded by National Science Foundation (NSF) Award No. DMS-1001012, ICB Award No. W911NF-09-0001 from the U.S. Army Research Office, NIBIB of the NIH under Awards No. R01-EB014877-01 and No. R01-GM113241, and (U.S.) Department of Energy (DOE) Award No. DE-SC0008975. The content of this paper is solely the responsibility of the authors and does not necessarily represent the official views of these agencies.
}}

\author{Brian Drawert$^{1}$
\and Michael Trogdon$^{2}$
\and Salman Toor$^{3,4}$
\and Linda Petzold$^{1,2}$
\and Andreas Hellander$^{4,5}$
}
\maketitle
\slugger{sisc}{xxxx}{xx}{x}{x--x}%

\footnotetext[1]{Dept. of Computer Science,
University of California, Santa Barbara,
Santa Barbara, CA, USA (\email{briandrawert@gmail.com}, \email{petzold@cs.ucsb.edu})}
\footnotetext[2]{Dept. of Mechanical Engineering,
University of California, Santa Barbara,
Santa Barbara, CA, USA (\email{mtrogdon@engineering.ucsb.edu})}
\footnotetext[3]{Dept. of Computer Science,
University of Helsinki,
Helsinki, Finland}
\footnotetext[4]{Dept. of Information Technology, Division of Scientific Computing,
Uppsala University,
Uppsala, Sweden (\email{salman.toor@it.uu.se}, \email{andreas.hellander@it.uu.se})}
\footnotetext[5]{Corresponding Author}

\begin{abstract}
Computational experiments using spatial stochastic simulations have led to important new biological insights, but they require specialized tools, a complex software stack, as well as large and scalable compute and data analysis resources due to the large computational cost associated with Monte Carlo computational workflows. The complexity of setting up and managing a large-scale distributed computation environment to support productive and reproducible modeling can be prohibitive for practitioners in systems biology. This results in a barrier to the adoption of spatial stochastic simulation tools, effectively limiting the type of biological questions addressed by quantitative modeling. 
In this paper, we present PyURDME, a new, user-friendly spatial modeling and simulation package, and \packagename, a cloud computing appliance for distributed simulation of stochastic reaction-diffusion models.
\packagename~is based on IPython and provides an interactive programming platform for development of sharable and reproducible distributed parallel computational experiments. 
\end{abstract}


\section{Introduction}
\label{sec:Introduction}

In computational systems biology, one of the main goals is to understand how intracellular regulatory networks function reliably in a noisy molecular environment. To that end, discrete stochastic mathematical modeling has emerged as a prominent tool.   
Stochastic simulation of well-mixed systems is now routinely used \cite{10659837,12237400,12183631,16179466}, and recently, spatial stochastic models have resulted in important scientific insights \cite{FaEl, lawson2013,Sturrock2013}, clearly demonstrating the potential as an analytic tool in the study of cellular control systems. Compared to less detailed models such as ordinary differential equations (ODE), well mixed discrete stochastic models or partial differential equations (PDE), spatial stochastic models are both more costly to simulate and more diffecult to formulate and set up. The large simulation cost of stochastic reaction-diffusion simulations has led to development of more efficient algorithms; an overview of theory and methods for discrete stochastic simulations can be found in \cite{GillHellPetz}.
Several software packages are publicly available, both for mesoscopic, discrete stochastic simulation \cite{mesoRD, steps, urdme} and microscopic particle tracking based on Brownian Dynamics (BD) \cite{smoldyn, eGFRD, MCell, ReaDDy}.  A recent overview of particle based simulators can be found in \cite{ScoUllNoe:2014}.

While efficient simulation methods are critical for well-resolved spatial models, practical modeling projects require the support provided by a software framework.  In the early stages of the model development process, 
there is typically no need for large compute resources. In later stages, 
computational experiments generate large numbers of independent stochastic realizations.
This is common to all applications that rely on Monte Carlo techniques. For spatial stochastic models, substantial computational and data handling facilities are required. A simulation framework that focuses on modeler productivity needs to accommodate both interactivity and visual feedback, as well as the possibility of large scale simulation and data handling. To be cost and resource efficient, it should also support dynamic scaling of compute and storage resources to accommodate the needs in different stages of the modeling process. 

Since most successful modeling projects involve a multidisciplinary team of researchers, it is important that models can be shared and understood by team members with different areas of expertise. Formats for model exchange based on static markup language descriptions such as the Systems Biology Markup Language (SBML) \cite{SBML} or Open Modeling EXchange format (OMEX) \cite{Bergmann2014} are useful to standardize descriptions of simple ODE and well-mixed stochastic models, but they fall short when it comes to complex spatial models. Recently, numerous developers of spatial simulation packages have taken another approach and provided application programming interfaces (APIs) for model specification in a scripting language \cite{pysb, steps, eGFRD}, with Python being a popular choice.  Our newly developed package PyURDME falls into this category.  We will show how PyURDME, being designed with the IPython suite in mind, can be used to program spatial stochastic models as highly interactive and sharable notebooks. 
In addition, we note that by providing a virtual cloud appliance, not only the models but also the computational experimental workflow including the computing environment becomes easily reproducible. 

In previous work, we have developed the URDME (Unstructured mesh Reaction-Diffusion Master Equation) framework for discrete stochastic simulation of biochemical reaction-diffusion systems \cite{urdme}. URDME was designed primarily as a traditional, native toolkit that combines MATLAB and COMSOL Multiphysics to form an interactive modeling and simulation environment. The  design of URDME has proven useful to support both methods development and modeling, but the framework has limitations when it comes to assisting large scale Monte Carlo computational experiments. URDME can be executed on clusters or grid resources \cite{urdme_bioinformatics2012}.
However, doing this typically requires computer science knowledge beyond that of the average practitioner, and access to High-Performance Computing (HPC) environments. 
This distracts users from the science problems addressed, and it acts as a barrier to scale up the computational experiments as needed for a consistent statistical analysis.  
Further, the computational experiment becomes hard to reproduce since the provenance relies on specific resources not accessible to third parties.  

Based on the above observations, we argue that the classical view of the scientific application (in our case PyURDME), as being separate from the compute, storage and data analysis tools employed, is restrictive. Enhanced modeling productivity and reproducibility would result if the computational infrastructure and the software stack were combined into a unified appliance. 
Hence, the aim of this work has been to develop a platform that: 

\begin{enumerate}
\item Allows interactive development of spatial stochastic models supported by basic visualization capabilities. 
\item Facilitates collaboration and reproducibility. 
\item Allows for convenient and efficient execution of common computational experiments, such as estimation of mean values, variances, and parameter sweeps. 
\item Is close-to-data and allows for flexible specification of custom post-processing. 
\item Allows for flexibility in the choice of computational infrastructure provider and dynamic scaling of computing resources.    
\item Requires no more than basic computer science knowledge to deploy and manage. 
\end{enumerate}

To meet all these requirements, we have developed \packagename, a cloud computing appliance that configures, builds and manages a virtual appliance for spatial stochastic modeling and simulation on public, private and hybrid clouds. 
By relying on cloud computing and its resource delivery model, the responsibility for handling the complex setup of the software stack is shifted from the users to the developers since we can prepare virtual machines that are pre-configured and ready to use. With support for the most common public clouds such as Amazon Elastic Compute Cloud (EC2) and HP Helion, we ensure high availability and scalability of computational resources. By supporting OpenStack, an open source cloud environment commonly used for private (in-house) cloud installations,  \packagename~brings the flexibility and tools of cloud computing to the user's own servers. Taking it one step further, \packagename~provides support for hybrid deployments in which private and public cloud resources can be combined, allowing the use of in-house resources  and bursting to public clouds during particularly compute-intensive phases of a modeling project. Interactivity is achieved by building on Interactive Python (IPython), in particular the web-based IPython Notebook project \cite{ipython, ragan2013collaborative}.

We demonstrate the potential of \packagename~to greatly assist computational experimentation in a case study of yeast polarization, and  evaluate its performance in parallel, distributed performance benchmarks. 
While the current computational engine is our newly developed Python package PyURDME, we believe that users as well as developers of other spatial simulation tools could benefit greatly from the delivery model proposed in our virtual platform. All components of the software presented here, as well as all models (and many more), are publicly available under open source licenses that permit unlimited redistribution for non-commercial purposes under the GPLv3 license at \url{https://github.com/MOLNs/MOLNs}.

\begin{figure}[htpb]
\begin{center}
\includegraphics[trim=0cm 0cm 0cm 0cm, width=\textwidth]{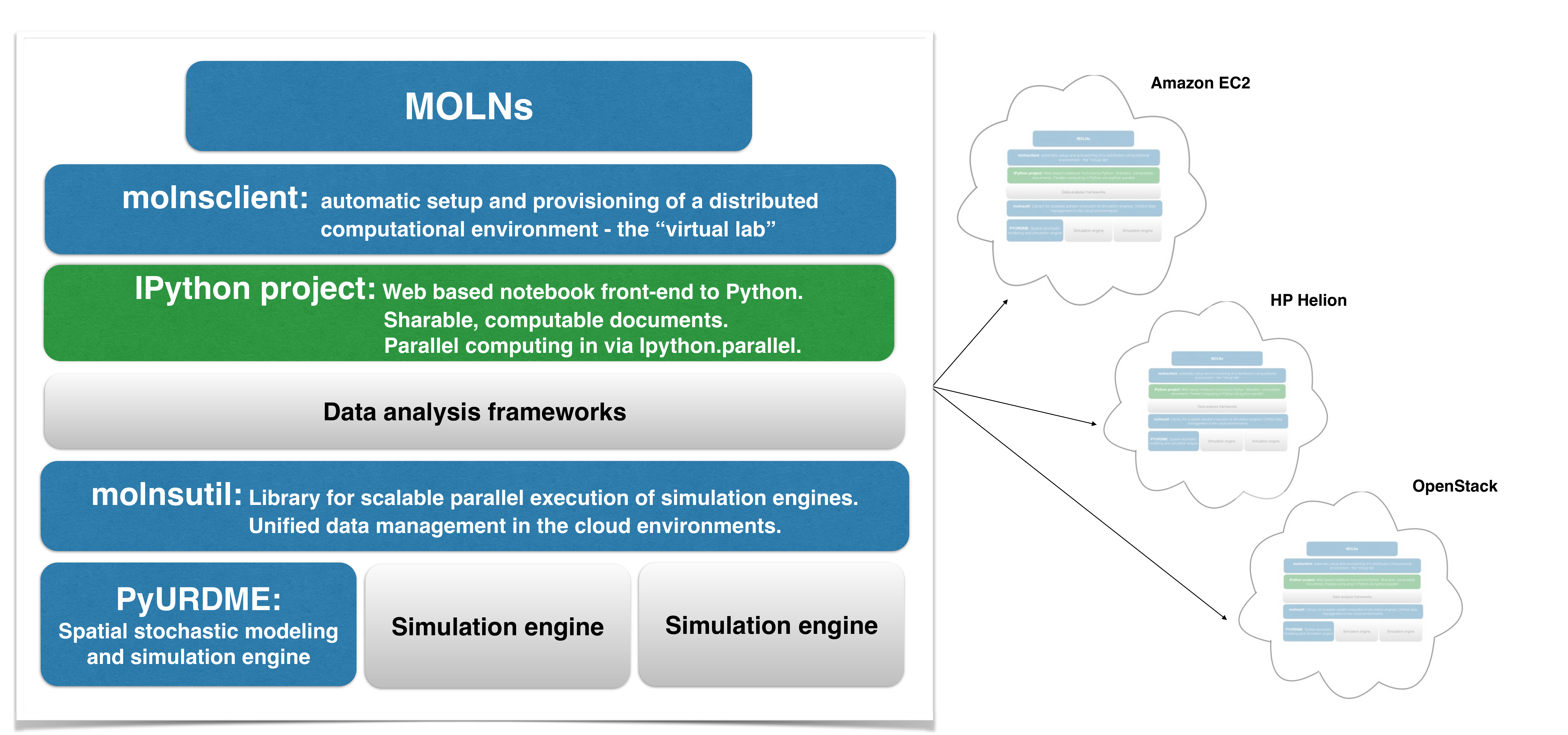}
\end{center}
\caption{{\bf MOLNs harnesses the power of cloud computing for biologists to use.}
Using MOLNs, biologists can take advantage of scalable cloud computing for compute-intensive computational experiments based on stochastic models of reaction-diffusion kinetics. Interactive modeling and scalable Monte Carlo experiments are provided through the use of the IPython Notebook and the newly developed libraries PyURDME and \emph{molnsutil}. Reproducibility of computational experiments requires more than sharing the model, or even the computational workflow that is used for the analysis. By creating a templated computational environment, MOLNs makes the entire "virtual lab" sharable, offering the flexibility to reproduce it in the infrastructure provider of choice, be that public cloud providers or in-house private clouds.  This ensures high availability and scalability.
Illustrated above are the main components of MOLNs, the newly developed ones are depicted in blue. Grey boxes illustrate the possibility to build on the proposed infrastructure and add additional data analysis tools to the virtual platform, such as Hadoop, Spark, or other simulation engines. 
}
\label{fig:box1}
\end{figure}

\section{Stochastic Simulation of Spatially Inhomogeneous Discrete Biochemical Systems}
Recent advances in biology have shown that proteins and genes often interact probabilistically. The resulting effects that arise from these stochastic dynamics differ significantly from traditional deterministic formulations, and have biologically significant ramifications. This has led to the development of discrete stochastic computational models of the biochemical pathways found in living organisms. These include spatial stochastic models, where the physical extent of the domain plays an important role. For mesoscopic models, similar to popular solution frameworks for partial differential equations (PDEs), the computational domain is discretized with a computational mesh, but unlike PDEs, the reaction-diffusion dynamics are modeled by a Markov process where diffusion and reactions are discrete stochastic events.
The dynamics of a spatially inhomogeneous stochastic system modeled by such a Markov process formalism are governed by the Reaction-Diffusion Master Equation (RDME)\cite{gardiner2004handbook}.

The RDME extends the classical well-mixed Markov process model \cite{SSA} to the spatial case by introducing a discretization of the domain into $K$ non-overlapping voxels. Molecules are point particles and the state of the system is the discrete number of molecules of each of the species in each of the voxels on Cartesian grids or unstructured triangular and tetrahedral meshes. 
The RDME is the forward Kolmogorov equation governing the time evolution of the probability density of the system.  
For brevity of notation, we let $p(\fatx,t) = p(\fatx,t|\fatx_0,t_0)$ for the probability that the system can be found in state $\fatx$ at time $t$, conditioned on the initial condition $\fatx_0$ at time $t_0$. For a general reaction-diffusion system, the RDME can be written as
\begin{align}
\label{eq:rdme}
\frac{\mathrm{d}}{\mathrm{dt}}p(\fatx, t) = 
&\sum_{i=1}^{K}
\sum_{r = 1}^{M}
a_{ir}(\fatx_{i \cdot}-\fatmu_{ir})p(\fatx_{1 \cdot},\ldots,\fatx_{i \cdot}-\fatmu_{ir},
\ldots,\fatx_{K \cdot}, t) \nonumber 
-\sum_{i=1}^{K}
\sum_{r = 1}^{M}
a_{ir}(\fatx_{i \cdot})p(\fatx, t)\\
&+\sum_{j=1}^{N} \sum_{i = 1}^{K} \sum_{k=1}^K d_{jik}(\fatx_{\cdot j}-\fatnu_{ijk})
p(\fatx_{\cdot 1},\ldots,\fatx_{\cdot j}-\fatnu_{ijk},
\ldots,\fatx_{\cdot N}, t) \nonumber\\
&-\sum_{j=1}^N\sum_{i=1}^{K}
\sum_{k = 1}^{K} d_{ijk}(\fatx_{\cdot j})p(\fatx, t),\nonumber\\
\vspace{-30pt}
\end{align}
\noindent
where $\fatx_{i\cdot}$ denotes the $i$-th row and $\fatx_{\cdot j}$ denotes the $j$-th column of the $K\times S$ state matrix $\fatx$ where $S$ is the number of chemical species. The functions $a_{ir}(\fatx_i)$ define the propensity functions of the $M$ chemical reactions, and $\fatmu_{ir}$ are stoichiometry vectors associated with the reactions. The propensity functions are defined such that  $a_{ir}(\fatx) \Delta t$ gives the probability that reaction $r$ occurs in a small time interval of length $\Delta t$. The stoichiometry vector  $\fatmu_{ir}$ defines the rules for how the state changes when reaction $r$ is executed. $d_{ijk}(\fatx_i)$ are propensities for the diffusion jump events, and $\fatnu_{ijk}$ are stoichiometry vectors for diffusion events. 
$\fatnu_{ijk}$ has only two non-zero entries, corresponding to the removal of one molecule of species $X_k$ in voxel $i$ and the addition of a molecule in voxel $j$.  The propensity functions for the diffusion jumps, $d_{ijk}$, are selected to provide a consistent and local discretization of the diffusion equation, or equivalently the Fokker-Planck equation for Brownian motion. 

The RDME is too high-dimensional to permit a direct solution. Instead, realizations of the stochastic process are sampled, using kinetic Monte Carlo algorithms similar to the Stochastic Simulation Algorithm (SSA)\cite{SSA} but optimized for reaction-diffusion systems. State-of-the-art algorithms such as the Next Subvolume Method (NSM)\cite{NSM} rely on priority queues and scale as $\mathcal{O}(\log_2(K))$ where $K$ is the number of voxels in the mesh. The computational cost of spatial stochastic simulation depends on the number of reaction and diffusion events that occur in a simulation, since exact kinetic Monte Carlo (KMC) methods sample every individual event. The number of diffusion events in the simulation scales as $O(h^{-2})$, where $h$ is a measure of the mesh resolution. This leads to stochastic stiffness, where diffusion events greatly outnumber reaction events for fine mesh resolutions. This has led to development of hybrid and multiscale methods to improve the situation. For an overview see \cite{GillHellPetz}. 

Despite the large computational cost, mesoscopic simulation with the RDME, when applicable, is typically orders of magnitude faster than alternatives such as reactive Brownian dynamics.  Individual realizations can be feasibly sampled for fairly complex models in complicated geometries on commodity computational resources such as laptops and workstations. However, since the models are stochastic, single realizations are not sufficient. Rather,  large ensembles of independent samples of the process need to be generated to form a basis for statistical analysis. 
Furthermore, key parameters of the biological process may be known only to an order of magnitude or two, thus necessitating an exploration of parameter space and/or parameter estimation. The need for an infrastructure to manage the computation and data has motivated the development of PyURDME and the MOLNs platform.

\section{Results}
\subsection{Construction of Spatial Stochastic Models with PyURDME}

\label{sec:pyurdme}

PyURDME (\url{www.pyurdme.org}) is a native Python module for development and simulation of spatial stochastic models of biochemical networks. It is loosely based on the URDME \cite{urdme} framework, in that it replicates the functionality of URDME's core and uses modified versions of the stochastic solvers. While URDME was designed as an interactive MATLAB package, using COMSOL Multiphysics for geometric modeling and meshing, PyURDME is a Python module providing an Object-Oriented API for model construction and execution of simulations. PyURDME relies only on open source software dependencies and uses FEniCS/Dolfin \cite{LoggMardalEtAl2012a} as a replacement for the facilities that COMSOL provided for URDME. 

\begin{figure}[htpb]
\begin{center}
\includegraphics[width=\columnwidth]{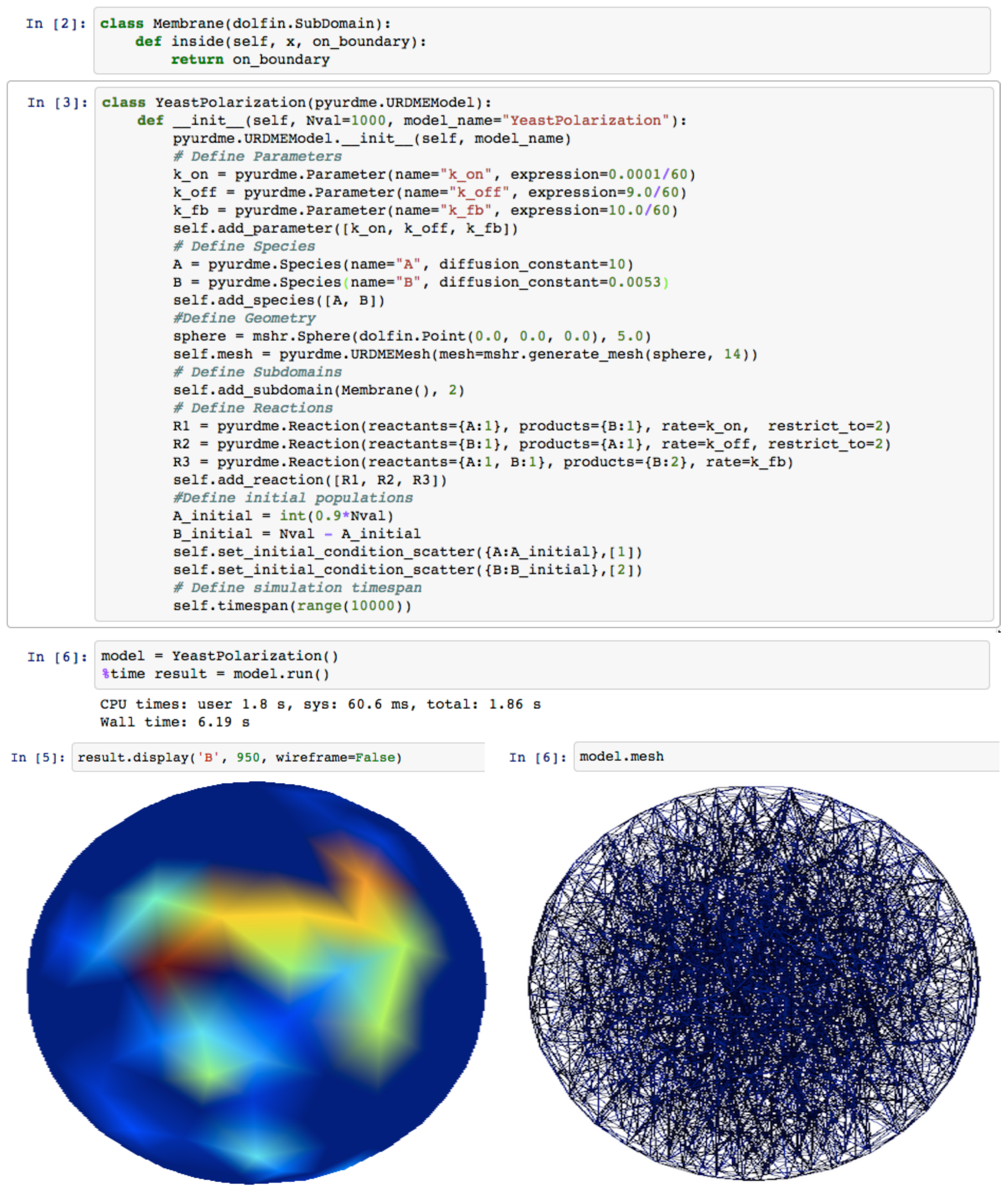}
\end{center}
\caption{Definition of the yeast polarization model, and examples of simulation and visualization that PyURDME and \packagename~provide within the IPython notebook interface.  This simple workflow demonstrates the usage of PyURDME.
}
\label{fig:fig1}
\end{figure}

Creating a model in PyURDME involves implementing a class that extends a base model, \emph{URDMEModel}, where information about chemical species, reaction rates, reactions and the geometry and mesh are specified in the constructor. This is a minimal amount of Python code that is easily readable and powerful enough to extend to more complex models quite intuitively. Then, spatial stochastic solvers, each based on a base-class \emph{URDMESolver}, can be instantiated from  a reference of the model. After executing simulations, results are encapsulated in an \emph{URDMEResult} object. The excerpt of an IPython notebook \cite{ipython} in Fig.~\ref{fig:fig1} illustrates specification and execution of a model of spontaneous polarization in yeast \cite{Altschuler2008}. We will use the development and analysis of this model as a case study later in this manuscript.
In the Supplementary Information, we provide in-depth explanations of the design and workings of the key classes \emph{URDMEModel}, \emph{URDMESolver} and \emph{URDMEResult}.

The \emph{URDMESolver} class provides an interface to spatial stochastic solvers.  The current core solver in PyURDME is a modified version of the NSM \cite{NSM} core solver in the URDME framework \cite{urdme}. It is implemented in C, and we follow the same execution mechanism as in \cite{urdme}.  Upon execution of the solver (e.g. the \emph{model.run()} command in Fig.~\ref{fig:fig1}), PyURDME uses the model specification encoded in \emph{YeastPolarization} to assemble the data structures needed by the core solver. It also generates a C file specifying the reaction propensity functions, and compiles a binary for execution of the specific model. The binary solver is then executed as a separate process.
The core solver execute the NSM method and generates a spatio-temporal time series data set which is written to the compressed binary file in the HDF5 format \cite{hdf5}. 

Though all of the functionality of PyURDME is available when using it as a native library on a local client (such as a user's laptop), we provide additional functionality to enhance the usability when integrated in \packagename. For example, simulation results can be visualized with a 3D rendering of the mesh or domain inline in IPython Notebook using the JavaScript library three.js \cite{ThreeJS}, (as illustrated in Fig.~\ref{fig:fig1}). Additionally, special attention has been paid to make the instances of \emph{URDMEModel}, \emph{URDMESolver} and \emph{URDMEResult} serializable.  This enables PyURDME to integrate with \emph{IPython.Parallel} library, the distributed computing facilities of IPython, and is an important property that prepares PyURDME for distributed computing. PyURDME models need not be developed in a tightly coupled manner on the \packagename platform, but a benefit of doing so is that it enables seamless integration with the development and visualization facilities, and allows the computational scientists to easily harness the computational power of the large scale distributed computational cloud computing environment.  

\subsection{The \packagename~Cloud Platform}

\begin{figure}[htpb]
\begin{center}
\includegraphics[width=.7\columnwidth]{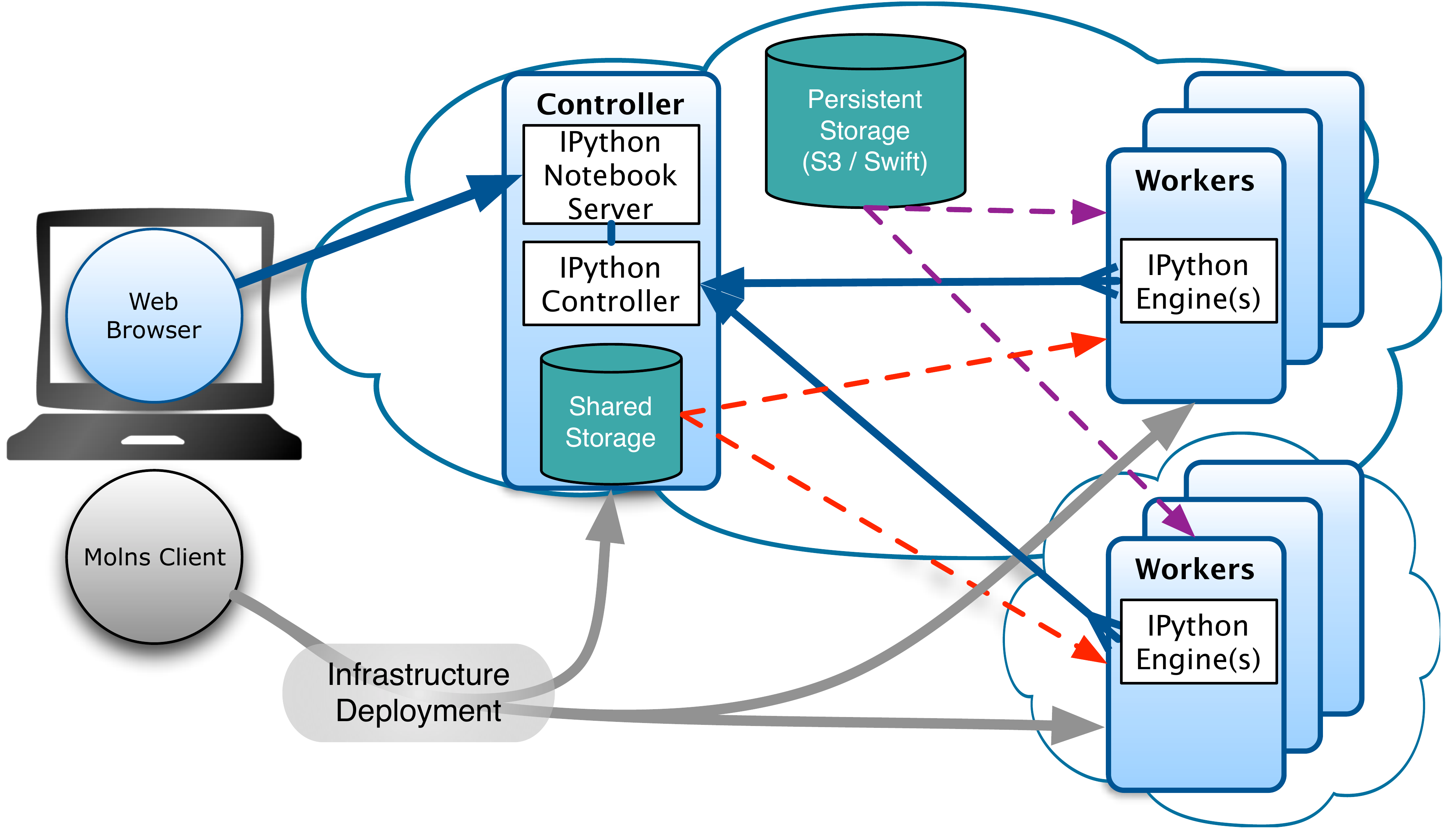}
\end{center}
\caption{\packagename~cluster architecture and communication.
Users interact with \packagename~in two ways; using the \emph{molnsclient} and a web browser.
The \emph{molnsclient} is used to create, start and stop a \packagename~cluster by provisioning \emph{Controllers} and \emph{Workers} on multiple clouds (gray arrows).  Once a cluster is active, the web browser is used to connect to the IPython notebook web-based interactive computational environment, which provides an interface to PyURDME for modeling and simulation, and to \emph{molnsutil} for distributed computational workflows which utilize the \emph{Workers} of the \packagename~cluster.  \emph{Molnsutil} distributes the computations via the IPython controller and IPython engines (blue arrows) and is able to store simulation data in either a transient shared storage (red arrows) or the persistent cloud object storage (i.e. Amazon S3 or OpenStack Swift, purple arrows).  
}
\label{fig:fig2}
\end{figure}

The  \packagename~cloud computing platform has three major components, as shown in Fig.~\ref{fig:fig2}. The first component is the IPython notebook web interface, which provides a widely used and familiar user interface for the \packagename~platform. The second component is the \emph{molnsclient}, a command line interface (CLI) which is responsible for the setup, provisioning, creation and termination of \packagename~clusters on private or public cloud computing infrastructure services.  The final component is the \emph{molnsutil} package which provides a high-level API for distributed simulation and post-processing of Monte-Carlo workflows with PyURDME.  Together, these components make up a powerful and easy to use tool for harnessing the computational power and high availability of cloud computing resources in an interactive, sharable and reproducible manner. %

\subsubsection{IPython Notebook Server}
\label{sec:ipython}
The first component of the \packagename~platform is an IPython notebook server. 
The IPython notebook is a web-based interactive computational environment where code execution, text, mathematics, plots and rich media can be combined into a single document. The main goal of the IPython project has been to provide interactive computing for scientists \cite{ipython}, and it has gained widespread use in the scientific community. IPython notebooks are ''computable documents'' and this makes them ideal to present easily reproducible and shareable scientific results \cite{ragan2013collaborative}. IPython Notebook was recently suggested in a Nature Editorial to be a promising tool for addressing the lack of reproducibility of computational biology results \cite{shen2014}. An example of the usage of PyURDME in such a notebook is shown in Fig \ref{fig:fig1}. 

While the notebooks contain the information needed to share and reproduce the model and the structure of the computational experiment, other important parts of the provenance of a computational experiment are the compute infrastructure and the software stack. For computational experiments, the software stack is often quite complex, and a notebook does not provide a way to set up an environment in which it can be executed. For spatial stochastic simulations, this is complicated further by the need for complex HPC infrastructure. This is addressed by \emph{molnsclient}.

\subsubsection{Molnsclient}
\label{sec:molns}
The second component of the \packagename~software is the \emph{molnsclient}, which is responsible for the infrastructure management of cloud computing resources.  It is a CLI for provisioning the \packagename~clusters, i.e. starting and terminating the virtual machine instances on the cloud computing service providers.  This is represented by the gray lines in Fig.~\ref{fig:fig2}.
The configuration of \emph{molnsclient} is organized into \emph{Providers}, \emph{Controllers} and \emph{Workers}. The CLI allows the user to configure and setup each of these objects.

A \emph{Provider} represents a cloud Infrastructure-as-a-Service (IaaS) provider, such as public cloud providers Amazon EC2\footnote{http://aws.amazon.com/ec2/} %
or HP Cloud\footnote{https://horizon.hpcloud.com/}, %
 or a private installation of cloud IaaS software such as OpenStack\footnote{http://www.openstack.org/} %
 or Eucalyptus \cite{eucalyptus}. To setup a \emph{Provider}, the user simply provides access credentials. Next, \emph{molnsclient} will automate the building of the virtual machine (VM) images.  This is done by starting a clean Ubuntu 14.04 seed VM. Then, using package management and source control programs, 
the set of packages necessary for \packagename~are loaded onto the image.  The image is then saved and used for all subsequent provisioning of virtual machines on this \emph{Provider}. 

A \emph{Controller} represents the head node of a \packagename~cluster. It is associated with a specific \emph{Provider}. It hosts the IPython notebook server interface, the parallel computing work queue (IPython parallel  controller), and hosts the \emph{SharedStorage} service.
If a \emph{Controller} VM has enough CPUs, one or more IPython parallel engines will be started on the node as well.
A \emph{Worker} represents one or more \emph{Worker} nodes and is associated with a \emph{Provider} and a \emph{Controller}.  It is not required that a \emph{Worker} has the same \emph{Provider} as its associated \emph{Controller}.  Indeed, starting \emph{Workers} on a different \emph{Provider} than the \emph{Controller} enables \packagename's heterogeneous cloud computing capability, see Fig.~\ref{fig:fig2}. \emph{Workers} host IPython parallel engines for parallel processing, typically one per CPU.  \emph{Controllers} and \emph{Workers} can be started independently and additional workers can be added and removed from a running cluster dynamically, though a \emph{Worker} can only be started if its associated \emph{Controller} is already running.

Together, the infrastructure set up by \emph{molnsclient} and the IPython framework provides an environment that allows interactive and efficient parallel computational experiments. However, the virtual cloud environment adds requirements for handling data not addressed by IPython. Also, directly using the IPython parallel APIs to script scalable Monte Carlo experiments requires some computer science expertise. Hence, there is a need to simplify for practitioners the setup and execution of typical experiments with the spatial stochastic solvers. These issue are addressed by the \emph{molnsutil} package.

\subsubsection{Automatic parallelization of systems biology workflows}
Providing access to massive computational resources is not sufficient to enable the wider community to utilize them.  Efficient use of parallel computing requires specialized training that is not common in the biological fields. Since we have designed MOLNs as a virtual platform, and thus control the whole chain from software stack to virtual compute infrastructure, building upon a state-of-the-art parallel architecture (IPython.parallel), we are able to implement a high level API that provides simple access to the parallel computational resources. From a modeler's perspective, this ensures that computational experiments can be scaled up to conduct proper statistical analysis or large scale parameter studies without having to deal with managing a distributed computing infrastructure. Instead, the modelers can spend their time on interactively developing and refining post-processing functions as simple Python scripts.

The role of \emph{molnsutil} is to bridge the gap between the underlying virtual infrastructure provisioned by \emph{molnsclient} and the modeler, providing easy to use abstractions for scripting Monte Carlo experiments in the IPython Notebook front-end.  IPython.parallel provides an API for distributed parallel computing that is easy to use for computational scientists. Using it, general parallel computing workflows can be implemented and executed in the MOLNs environment. In \emph{molnsutil}, we have used this API to provide high-level access to the two most common computational workflows with PyURDME: the generation of large ensembles of realizations and global parameter sweeps. We also address the question of data management in the cloud as this issue is out of the scope of the IPython environment. The \emph{molnsutil} library provides an easy-to-use API to store and manage data in MOLNs.

\begin{figure}[htpb]
\begin{center}
\includegraphics[width=.65\columnwidth]{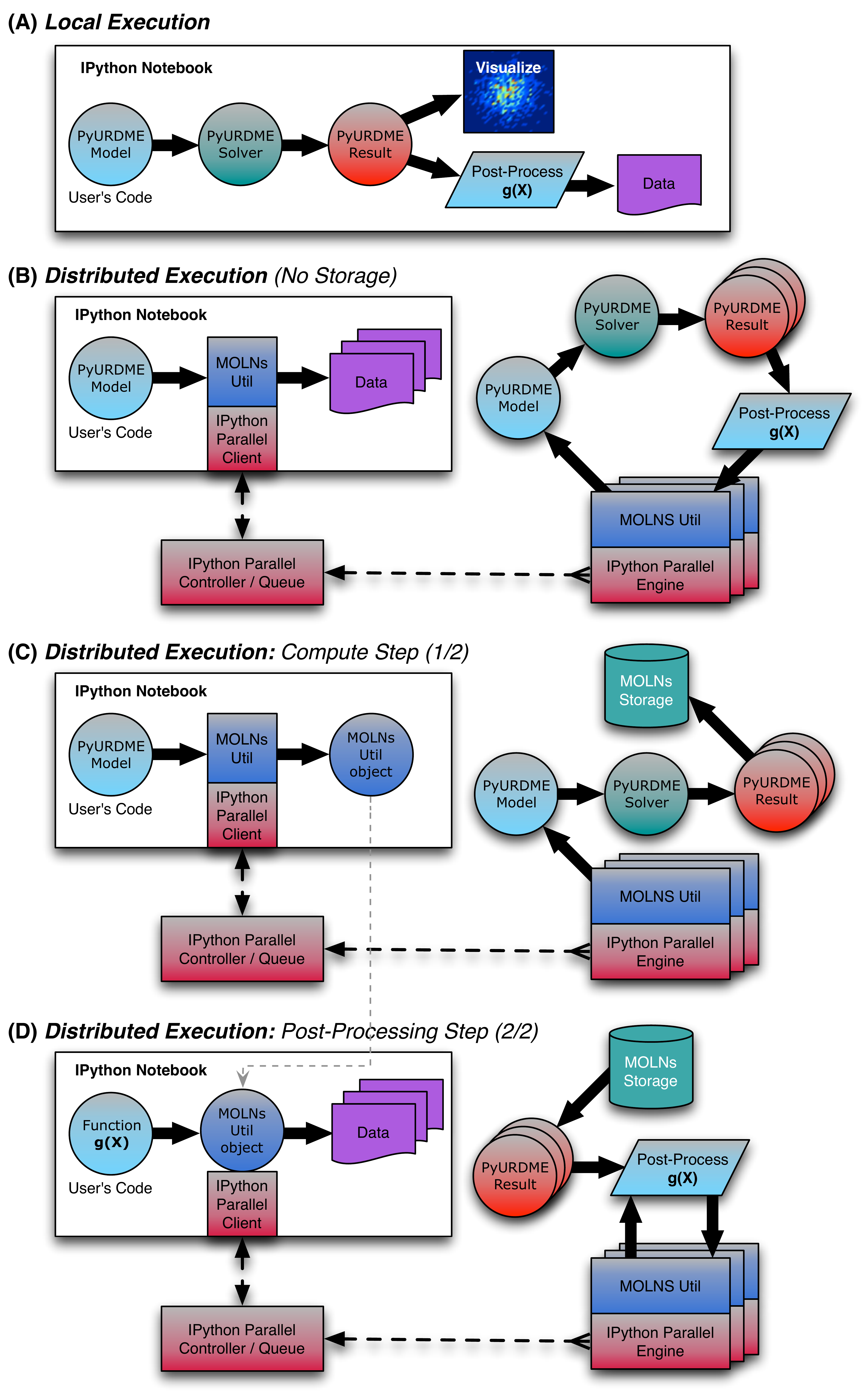}
\end{center}
\caption{\packagename~workflows.
{\bf (A)} Basic workflow executed within the IPython notebook.  The user develops a biological model, and the model is executed by the solver to produce a result object. The results are either visualized using functionality in PyURDME, or passed to a user-defined post-processing function $g(x)$. This local simulation workflow does not require \emph{molnsutil}, and can hence be developed locally without cloud resources. 
{\bf (B)} Distributed computational workflow. The user develops a biological model and post-processing function, passes them to \emph{molnsutil} which arranges the distributed execution into tasks and enacts it using IPython parallel. Each task executes the model to produce one or more result object which are processed by the user supplied $g(x)$.  The resulting data is aggregated and returned to the user's IPython notebook session.
{\bf (C)} In many cases it is advantageous to separate the generation of the result objects from the post-processing.  This shows the distributed workflow of generating the results and storing them in the integrated storage services so that subsequent runs of the post-processing analysis scripts {\bf (D)} can be done multiple times, allowing interactive development and refinement of these scripts.
}
\label{fig:workflows}
\end{figure}

\subsubsection{Cluster and cloud storage API}

The storage API mirrors the storage layers in the infrastructure, see Figure \ref{fig:fig2} and Table \ref{tab:storage}. We define three API-compatible storage classes: \emph{LocalStorage}, \emph{SharedStorage}, and \emph{PersistentStorage}, where the first enables writing and reading of files to the local ephemeral disks of the individual compute nodes, the second uses the cluster-wide network storage and the third uses the Object Store of the underlying cloud provider. They all fulfill separate needs; \emph{LocalStorage} is used for caching files near compute engines and has the smallest I/O overhead, but adds complexity for the developer in dealing with failures that lead to data loss. This storage mode is mainly used internally in \emph{molnsutil} for optimization purposes. \emph{SharedStorage} provides a non-persistent global storage area that all compute engines can access, making  the computations more robust to failing workers. Using \emph{SharedStorage} does not incur any additional cost beyond the cost for the deployed cluster instances\footnote{If all VMs are within the same availability zone}. \emph{PersistentStorage} also provides global access to objects, but in addition makes them persistently available outside the scope of the deployed cluster, and visible to other applications (if they share credentials to  access the storage buckets). {PersistentStorage} is hence ideal for simulation data that needs to be shared or stored for long periods of time. In public clouds, using \emph{PersistentStorage} incurs extra cost both for storing the objects and for accessing them. As long as the cluster is deployed in a sensible manner, current cost models in the supported clouds permit free network transfer from the object store to the compute engines. In addition to storage abstractions, \emph{molnsutil} contains parallel implementations of two important Monte Carlo computational workflows. 

\begin{table*}[t]
\begin{tabular}{|l|l|l|l|}
\hline
{\bf Type} & {\bf Advantages} & {\bf Disadvantages}\\
\hline
SharedStorage  &  No additional cost for read/write & Total storage limited to Controller disk size\\
& Fastest throughput for small clusters & Non-redundant storage\\
& No management of remote data & Throughput limited on large clusters \\
\hline
PersistentStorage   & Persistent data  & Storage and access incur cost\\
  &  Designed for extreme scalability  & \\
\hline
LocalStorage  & Best data locality & Non-robust to worker failure \\
  &   High I/O throughput & Increased complexity for developer\\
\hline
No-Storage  &  Best parallel scaling & Data must be recomputed for analysis\\
& No cost for data storage	& \\
\hline
\end{tabular}
\caption{Comparison of storage types available to \packagename~distributed workflows.}
\label{tab:storage}
\end{table*}

\subsubsection{Ensemble statistics}
\label{sec:statistics}
A frequently occurring scenario in computational experiments with spatial stochastic solvers is the generation of very large ensembles of independent realizations from the same model, followed by a statistical analysis based on the ensemble data. Often, a post-processing function is used to translate the detailed state information $\mathbf{X}$ into information directly related to the biological question being addressed. Hence, this function $g(\mathbf{X})$ is provided by the user. The most common statistics are the mean and the variance. The mean of $g(\mathbf{X})$, $E[g(\mathbf{X})]$, can be computed as 
$E[g(\mathbf{X})] = {\frac{1}{K}}\sum_{k=1}^{K} g(\mathbf{X}_k) $
where $K$ is the number of realizations in the ensemble, typically a large number. The variance is given by 
$V[g(\mathbf{X})] = {\frac{1}{K-1}}\sum_{k=1}^{K} (g(\mathbf{X}_k) - E[g(\mathbf{X})])^2$
and a $95\%$ confidence interval for the mean is then given by 
$E[g(\mathbf{X})] \pm 1.96\sqrt{(V[g(\mathbf{X})])/K}$.

In \emph{molnsutil}, this fundamental computation is implemented as part of a \emph{DistributedEnsemble} class. When generating a distributed ensemble, a \emph{URDMESolver} instance is created from a \emph{URDMEModel} class on each of the workers. The stochastic solver is then run (in parallel) independently to create the $K$ realizations. Each realization is represented by a \emph{URDMEResult} object. Hence, the $K$ \emph{URDMEResult} objects contain the $\mathbf{X}_k$ variables in the equations above. To compute the ensemble statistics we apply the post-processing function to all the results and aggregate them by summation. 

In Fig.~\ref{fig:workflows} we further distinguish two main variants of the execution of this fundamental workflow. First, where we do not store the intermediary data and instead directly pass it to the next part of the computation, only storing the final post-processed result (B). Second, where we in a first pass generate the ensemble data and store the intermediary result (the \emph{URDMEResult} objects) (C), and then, in a second pass, apply the post-processing routines and compute statistics (D). Both these cases are common in practical modeling projects. In early stages of a project, where the post-processing functions are being developed, one tends to favor storing the ensemble data and then interactively and dynamically analyzing it while developing the code for the post-processing analysis. Thus, the lifetime of the data may be hours to days and typically follow the lifetime of the compute cluster (making the use of \emph{SharedStorage} ideal). Later in the project when production runs are conducted, the generation of the ensemble data can require significant CPU time, and one may want to store the simulation data during the lifetime of the project (months to years) for re-analysis, reproducibility or sharing with another modeler. In this case, the lifetime of the data can be much longer than the lifetime of the cluster resources (making the use of \emph{PersistentStorage} ideal). In other situations, the stochastic simulations may run fast while the size of the output data set is large. In those cases, it may be preferable to simply recompute the ensemble data in every pass of an analysis step since the cost of re-computation is smaller than the cost of storage. 

Fig.~\ref{fig:distributedensemble} shows an excerpt from a \packagename~notebook illustrating how the above workflows are executed using \emph{molnsutil}. The user writes the post-processing function showed in cell \emph{In [7]}, and then in cell \emph{In [8]} creates an instance of \emph{DistributedEnsemble} and generates 200 realizations of the model, corresponding to the workflow in Fig.~\ref{fig:workflows}C.  Then, cell \emph{In [10]} executes the post-processing workflow in  Fig.~\ref{fig:workflows}D. Note that in order to change the analysis in a subsequent step, it is only necessary to modify the function $g$ in cell \emph{In [7]} and re-executing cell \emph{In [10]}. This gives the modeler the ability to interactively and efficiently develop the analysis functions. While it is not possible or desired to abstract away the user input for the analysis function, as this is where the biology question gets addressed, we have made efforts to abstract away the details of the numerics by encapsulating the data in the \emph{URDMEResult} object and exposing it through simple API calls. 
 
Table \ref{tab:storage} summarizes how the storage classes in \emph{molnsutil} maps to the different variants of the workflows. When creating a DistributedEnsemble, the default is to use \emph{SharedStorage}, but the user can switch to \emph{PersistentStorage} via a single argument to $add\_realizations.$  \emph{LocalStorage} is used internally to optimize repeated post-processing runs by explicitly caching data close to compute nodes.

\subsubsection{Parameter sweeps}
\label{sec:parametersweeps}
In most biological models, there is considerable uncertainty in many of the involved parameters. Experimentally determined reaction rate constants, diffusion constants and initial data are often known to low precision or not known at all. In some cases, phenomenological, or macroscopic outputs of the system are available from experiments, frequently in terms of fluorescence image data or coarse-grained time series data for the total number (or total fluorescence intensity) of some of the species. Hence, parameter sweeps are prevelant in modeling projects. Early in a modeling project, they are typically used for parameter estimation, i.e. finding values of the experimentally undetermined parameters that give rise to the experimentally observed phenomenological data. Such brute force parameter estimation may seem like a crude approach, but more sophisticated techniques based on e.g. parameter sensitivity have yet to be theoretically developed and made computationally feasible for mesoscopic spatial stochastic simulations. Later in a modeling project, when some hypothesis or observation has been made, it is typically necessary to conduct parameter sweeps to study the robustness of this observation to variations in the input data. We also note that studying the robustness of gene regulatory networks in a noisy environment has been a common theme in the systems biology literature \cite{12237400, Sturrock2013, lawson2013}.

\begin{figure}[htpb]
\begin{center}
\includegraphics[width=.75\columnwidth]{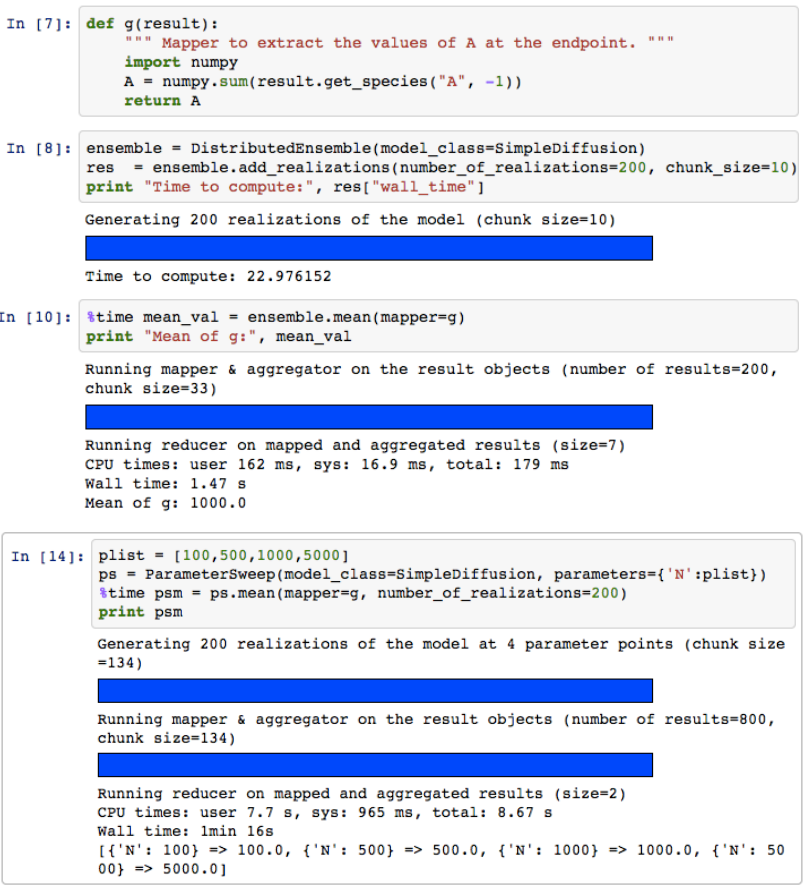}
\end{center}
\caption{Example usage of the DistributedEnsemble and ParameterSweep classes in \emph{molnsutil} inside an IPython notebook.  The blue bars are animated progress bars. }
\label{fig:distributedensemble}
\end{figure}

From a computational point of view, a parameter sweep can be thought of as generating a collection of ensembles, one for each parameter point being explored. Since the number of parameter points in a multi-dimensional parameter space grows very quickly with the number of parameters included in the sweep, the amount of compute time and storage needed for a parameter sweep can be very large, even if relatively small ensembles are generated for each parameter point. The same tradeoffs with respect to storing the ensemble trajectory data as discussed above for a single ensemble applies also to parameter sweeps, but due to the massive amounts of data that is generated even for a moderately large parameter sweep, it will likely be more common to use the execution model where the time course simulation data for each (parameter, ensemble)-pair is not stored. In those cases, the evaluated output metrics for each parameter point will be stored for further analysis and visualization.  

The last cell in Fig \ref{fig:distributedensemble} shows how to execute a parameter sweep in \packagename.  The user simply provides a mapping between any named argument to the constructor in the \emph{URDMEModel} class definition and a value range. The \emph{molnsutil} package then executes the parallel workflow and returns the result.

\subsection{Case Study: Interactive model development, simulation and analysis of yeast polarization}
\label{sec:CaseStudy}

To illustrate the capabilities of \packagename~we created, implemented, and analyzed a model of spontaneous yeast polarization. The cell signaling system in \emph{Saccharomyces cerevisiae} is an ideal candidate to test \packagename~because it is a well studied, yet not fully understood system in which polarization is critical to the cell cycle. In this case study we describe how we developed our model, parameter sweeps, and post-processing analysis using \packagename, and reproduced conclusions from the literature. 

Yeast cells exist in both haploid and diploid forms, both of which can reproduce through mitosis (i.e. budding). The haploid cells exist in two different types which can mate with each other to form a diploid cell. In both cases, polarization of proteins into a cap within the cell is crucial to establish the point of budding or the point of the mating projection.  Cdc42 is a critical protein to the establishment and regulation of polarity \cite{pruyne2000}. Though many models exist, varying in range of mathematical complexity and physical relevance, we focus on a relatively simple model presented in \cite{Altschuler2008, Altschuler2011} that makes use of a minimal positive feedback circuit.

\subsubsection{Model specification}

The yeast cell is modeled as a sphere with a membrane on the surface of the sphere. The model has three reactions between two species: cytosolic Cdc42 is allowed to spontaneously attach to the membrane with rate $k_{on}$ (Eq.~\ref{eq:1}), membrane-bound Cdc42 can likewise spontaneously detach with  rate $k_{off}$ (Eq.~\ref{eq:2}), and finally membrane-bound Cdc42 can recruit cytosolic Cdc42 to the membrane at rate $k_{fb}$ to close the positive feedback loop (Eq.~\ref{eq:3}).
\begin{align}
 Cdc42_c &\xrightarrow{k_{on}} Cdc42_m \label{eq:1} \\
 Cdc42_m &\xrightarrow{k_{off}} Cdc42_c \label{eq:2} \\
 \qquad \qquad \qquad \qquad \qquad \qquad Cdc42_c + Cdc42_m &\xrightarrow{k_{fb}} 2 Cdc42_m \label{eq:3}
\end{align}
The cytosolic and membrane bound species can diffuse at rates $D_{cyt}$ and $D_{mem}$ respectively (the diffusion of the membrane-bound Cdc42 being restricted to the membrane). The geometry, model definition, post-processing, and visualization are handled completely within the \packagename~environment.

The definition of the yeast polarization model is a Python class that extends the \emph{URDMEModel} class.
First, the model parameters and species were defined through \emph{add\_parameter} and \emph{add\_species} functions with the expressions and names for model parameters and species (these and all other commands referenced can be seen explicitly in the example code provided in Fig.~\ref{fig:fig1}).
Next, we defined the geometry of the cell using the built-in functionality of the FEniCS/DOLFIN \cite{LoggMardalEtAl2012a, dolfin} constructive solid geometry (CSG) sphere object.  
The membrane subdomain is defined by a custom class that marks all voxels on the surface of the sphere, and is then added to the model object via the \emph{add\_subdomain} function.
Next the reactions are defined by specifying the reactants, products, rate parameters, and subdomains that reactions and species are restricted to. In this example problem all reactions are mass action, but PyURDME also has the capability to take custom propensity functions for a reaction, such as Michaelis-Menten. The reactions are added to the PyURDME model object via the \emph{add\_reaction} function.
The last step in the model definition is to provide initial conditions and information about the simulation time span. Here, initial conditions were specified to be a random scattering of 10\% of molecules on the membrane and the rest scattered through the cytosol. 
Although this example is intended to be simple, the design of PyURDME enables easy extension of these modeling definition techniques to much more complex systems. All code and parameter values for this model can be found in the attached example files in the Supplementary Information. 

\subsubsection{Model execution, parameter sweep and post-processing}
Once we have completed the model definition, we execute the simulation within the same IPython notebook with one \emph{run()} command. After model execution, the post-processing capabilities of \packagename~can be utilized. Having all model parameters, species, geometry, subdomains, and reactions organized within one easily accessible PyURDME model object simplifies the development of  post-processing analysis scripts. All of the post-processing and data visualization take place right in the same IPython notebook in which model definition and execution occurred. All computation is performed in the cloud and the users interact via a web browser connected to the IPython notebook interface. In particular, interactive 3D plots of results are rendered in the web browser.

The IPython notebook contains the code that generates plots and the interactive plots themselves within one editable, easily transferable document, which provides the \packagename~user a unique modeling experience that significantly eases the development process. A result can be visualized right along with the code that generated it and any errors or changes that need to be made will be readily apparent. For this particular example it was of interest to monitor the polarization activity on the membrane. The previous implementations of this positive feedback model \cite{Altschuler2011} made explicit predictions of a density-dependent switch that drives stochastic clustering and polarization (although the physical relevance of this behavior has more recently come into question  \cite{Frei1}). 

To determine whether the density-dependent switch behavior was in fact observed, we varied the total number of Cdc42 signaling molecules while keeping the volume constant and investigated the polarization behavior. The interactivity of \packagename~allowed useful data to be easily stored and analyzed, which in turn led to the development of metrics quantifying polarization over time.

\subsubsection{Result interpretation and case-study summary}
One result that the design of \packagename~facilitated was to define a polarization metric that tracks the clustering behavior of the membrane molecules over time. The number of molecules at each voxel is stored for every time point in the PyURDME result object. This allowed the number of membrane molecules to be plotted over time, and once some dynamic equilibrium state is reached, the clustering can be investigated. Here, polarization at any given time was defined by a region making up 10\% of the membrane surface area containing more than 50\% of the membrane molecules.
This metric could be monitored and plotted for each number of signaling molecules to try to discern a qualitative density-dependent switch behavior for polarization. 

\begin{figure}[htpb]
\begin{center}
\includegraphics[width=.55\columnwidth]{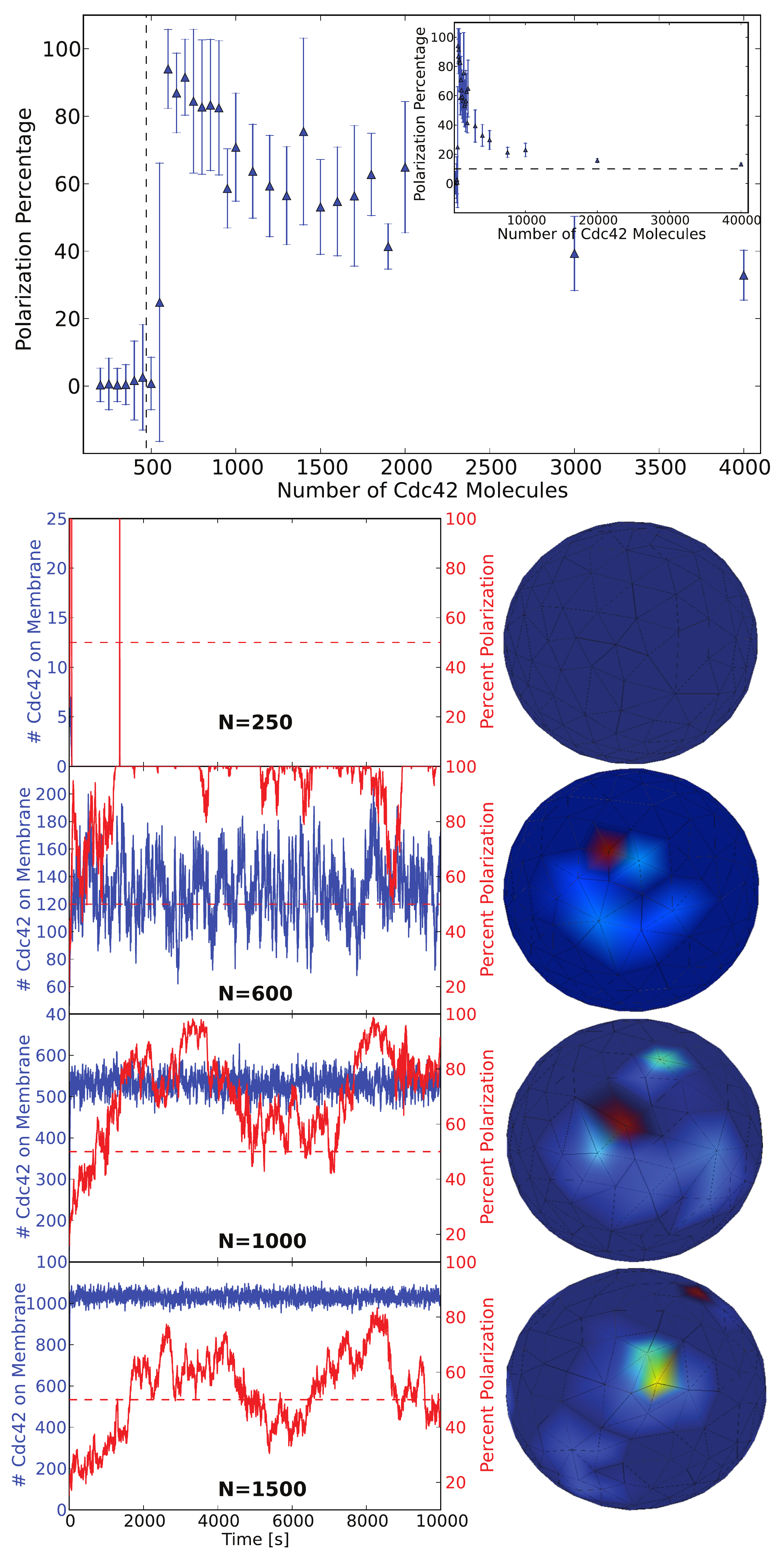}
\end{center}
\caption{
The results of a parameter sweep over the number of Cdc42 signaling molecules, N, with volume held constant, performed in parallel. Each model with a given parameter value of N was run to time 10,000 seconds. Plotted (top) is the time average of the maximum percent of Cdc42 molecules found in any region corresponding to ten percent surface area on the cell membrane for each N value, with error bars depicting the standard deviation. The dotted line represents the theoretical switch location calculated from \cite{Altschuler2011}. The model captures both the theoretical density dependent switch behavior and the asymptotic decrease to a homogeneous distribution, which corresponds on average to a maximum of ten percent of molecules in any ten percent region on the membrane. Plotted (bottom) is explicit polarization percentage and number of Cdc42 molecules versus time for various values of N along with a characteristic 3D visualization for each. It is important to note that at N=250 there is no membrane bound Cdc42 as it all remains in the cytoplasm throughout the simulation, which will always be the case below the switch value.}
\label{fig:polarization_percent}
\end{figure}

A parameter sweep was run in parallel for a range of Cdc42 molecule counts. Each parameter point was analyzed using a custom post-processing function to calculate polarization percent versus time. In this case it was not necessary to store the large amounts of data from the intermediary simulations, but rather return only the output of the post-processing function for each parameter point, thus we used the \emph{No-Storage} method in \emph{molnsutil}.  Plots of polarization percent versus time along with the total number of membrane bound Cdc42 molecules versus time for various numbers of total Cdc42 molecules can be seen in Fig.~\ref{fig:polarization_percent}. Based on the predictions of \cite{Altschuler2011}, there should be a critical range for polarization. This range is from a lower critical number of molecules necessary to facilitate polarization to an upper number above which molecules essentially become homogeneous on the membrane (i.e. not polarized). In Fig.~\ref{fig:polarization_percent} the time average of the maximum polarization percent is plotted for each Cdc42 molecule count, with error bars corresponding to the standard deviation. As can be seen in Fig.~\ref{fig:polarization_percent}, there is in fact a density-dependent switch behavior in the model. Below the theoretical critical value calculated from \cite{Altschuler2011} (around 500 molecules for this model) the molecules are in a homogeneous off state, meaning all of the molecules stay in the cytosol. There is an abrupt switch to a high percent polarization above the critical value. As the number of molecules is increased further, they asymptotically approach a homogeneous distribution on the membrane, as predicted by \cite{Altschuler2011}.

This case study illustrates the power and ease with which \packagename~users can define and analyze biologically relevant models. Having a coding environment for model and post-processing development and the interactive visualization of results side by side in one self contained document with all computation taking place in the cloud makes for a smooth development experience. Also the ability to perform large scale parameter sweeps efficiently in the cloud and to effectively organize the results is crucial for any modeling task.  

\subsection{Parallel computing performance}
\label{sec:benchmark}
Since \packagename~builds on the IPython suite, it inherits a design focused on interactive parallel computing and dynamic code serialization (enabling the interactivity in the development of the post-processing routines), hence programmer productivity and flexibility are areas were \packagename~can be expected to excel. As we have seen, this is enforced by the design of PyURDME. However, parallel performance and scalability are also important factors to consider since they map directly to cost in public cloud environments.
Here, we study the performance for our most fundamental computational workflow: generation of a distributed ensemble and subsequent post-processing by computing the ensemble mean for a given function. We examine the performance in three different clouds: MIST, a privately managed OpenStack Havanna deployment,  Amazon EC2 and HP Helion public clouds. Finally, we benchmark the system for a hybrid deployment using the HP and EC2 providers. Details regarding the IaaS providers and instance types can be found in the Supplementary Information.

Fig.~\ref{fig:benchmarks} shows strong and weak scaling experiments when executing the workflows B--D detailed in Fig.~\ref{fig:workflows}. Strong scaling shows the execution time for a fixed problem size, here computing an ensemble with $10^2$ realizations, with increasing number of computational engines. We start with a relatively small number of realizations to highlight the impacts of system latency on how much the simulation time can be reduced for a given problem by adding more workers to the \packagename~cluster. Weak scaling on the other hand shows the total execution time when both the number of engines and the problem size are increased proportionally so that the total work per engine stays constant. This benchmark shows how well the system lets you increase the problem size by scaling out the compute resources, and the ideal outcome is a constant execution time independent of problem size. In reality, the execution time will never be perfectly constant due to the possibility of exhausting common resources such as disk I/O throughput or network bandwidth (in the case of storing simulation data) or due to scheduling overheads as the number of tasks and workers become numerous. Since these particular workflows map well to the MapReduce programming model, as do many simple Monte Carlo experiments, we will also compare the performance of the \packagename~implementation to a reference implementation using Hadoop streaming on a virtual Hadoop cluster deployed over the same physical resources in our private cloud, MIST.  Details of the Hadoop implementation can be found in the Supplementary Information. 

It is not our objective to compare the performance of the different cloud providers in absolute numbers since the underlying hardware differs, although we chose instance types that are as closely corresponding to each other as possible (details can be found in the Supplementary Information). Rather, we are interested in the scaling properties which we find to be similar for all cloud providers as can be seen in Fig.~\ref{fig:benchmarks}. For strong scaling, irrespective of storage mode, we initially see a rapid reduction in simulation time and a saturation for larger number of engines. This is expected due to the total overhead of the system that sets a fundamental limit on the possible speedup. The total simulation time at saturation is less than 20 seconds. For weak scaling, the \emph{SharedStorage} method is faster than using \emph{PersistentStorage} for a smaller number of nodes, however as the number of workers increases the PersistentStorage is scaled better. We find the crossover point for the performance of these two modes to be approximately 5 nodes. We also note that for the public clouds (in particular for EC2), the \emph{PersistentStorage} backend results in nearly perfect weak scaling, as the scaling curves parallel the \emph{No-Storage} curves. This result is expected since Amazon S3 object storage service used by the PersistentStorage backend is designed to handle large throughput. In the private cloud MIST, the OpenStack Swift object store uses a single proxy-server, which limits the load-balancing capabilities and as a result we see a linear scaling of computational time with respect to the total number of requests. In contrast, the \emph{SharedStorage} shows a limited capability to scale-out (add nodes) computations as the computational time increases sharply as the problem size becomes large.  This is a result of saturation of the I/O read and write throughput used by the \emph{SharedStorage} backend on the controller node. In terms of absolute numbers, the EC2 provider outperforms both HP or MIST cloud providers. One possible explanation for this would be the fact that the EC2 instances are equipped with SSD-disks which allow for faster I/O throughput.

For comparison, we performed these benchmarks using the widely used Apache Hadoop\footnote{http://hadoop.apache.org/} distributed computing software system.  Hadoop MapReduce implements parallel processing of data sets that are typically stored in the Hadoop Distributed File System (HDFS).  We performed the benchmarks on our private cloud MIST, and found that Hadoop with HDFS is slower than \packagename~for all cases. For weak scaling, Hadoop without storage is very close to \packagename~with \emph{No-Storage}, which is expected since the task size is large and system latency has little impact on the computational time. 

In addition to benchmarks on single cloud providers, we performed benchmarks on hybrid deployments where the controller node is on one cloud and all of the workers are on a separate cloud provider.  Hybrid deployments become useful when a user have exhausted their quota in one cloud and want to add more workers in a different cloud, or if they have access to a private cloud but want to burst out to a public cloud for meeting peak loads. For hybrid \packagename~deployments, the performance of computations using \emph{SharedStorage} scales badly due to the network latency for workers writing to the shared disk on the controller in a different cloud provider, to the point that its use cannot be recommended in a hybrid deployment (lower two panels). As can be seen, with \emph{PersistentStorage} or \emph{No-Storage}, a user can benefit from adding workers in a different cloud. It should be noted however, that the cost of using the \emph{PersistentStorage} in this case will be much higher than in the pure cloud environments since data is moved between cloud providers.

In conclusion, these benchmarks show that \packagename~is not only capable of providing a flexible programming environment for computational experiments, but also a scalable and efficient execution environment, also in comparison with less easy-to-use and less flexible systems such as the  industry-standard Apache Hadoop. 
We estimate total cost to run this suite of benchmarks was \$158 for the HP cloud provider and \$50 for the EC2 cloud provider (December 2014 prices). This estimate is based on the monthly billing statement, details can be found in the Supplementary Information.

\begin{figure}[htpb]
\begin{center}
\includegraphics[width=.75\columnwidth]{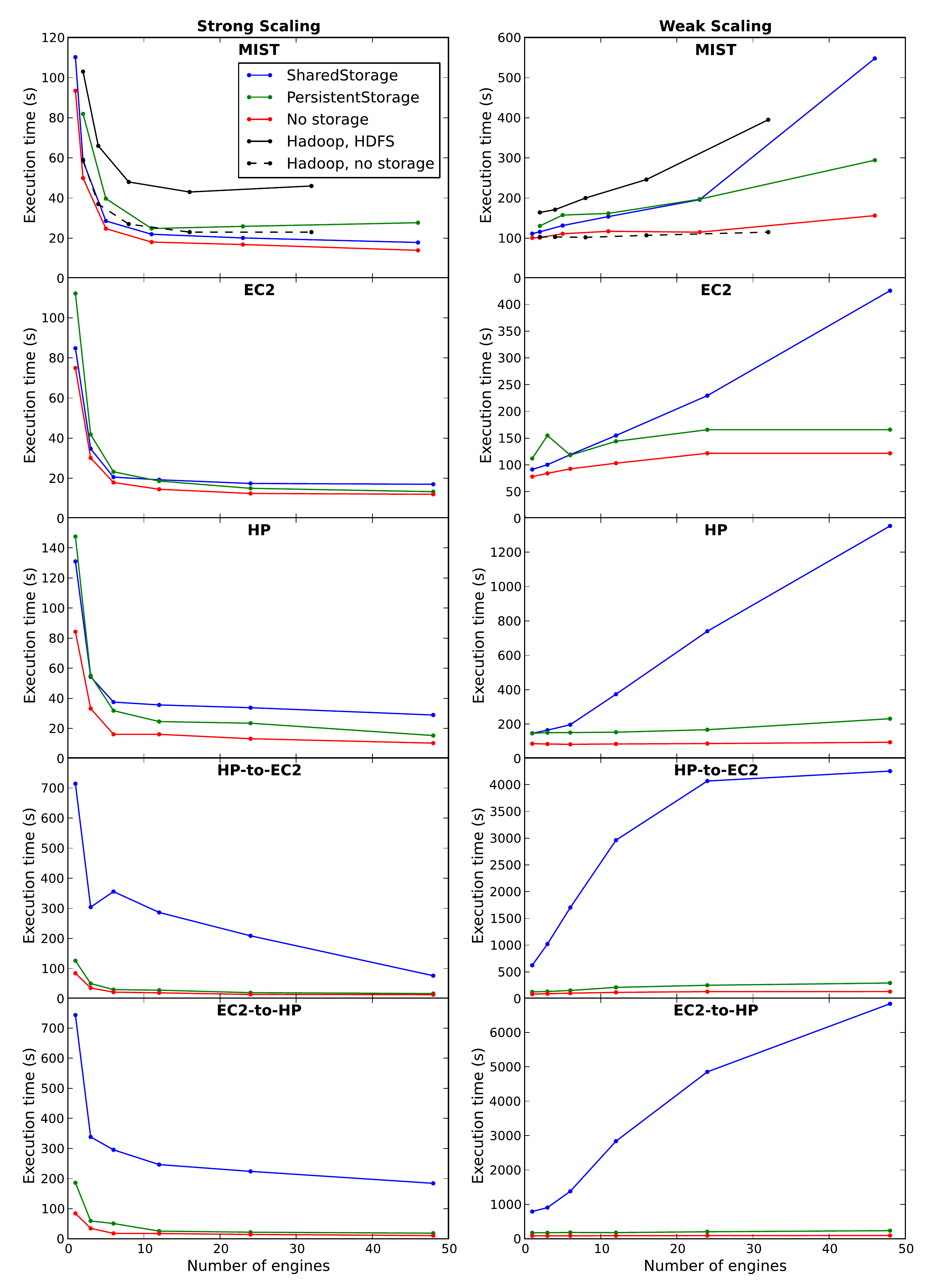}
\end{center}
\caption{
Benchmarks of \packagename~for a computation and analysis workflow of a PyURDME distributed ensemble.  The left column shows strong scaling tests which demonstrate parallel efficiency; a constant number of jobs (100) executed on a varying number of engines.  The right column shows weak scaling tests which demonstrate efficiency of scaling up the problem size; a constant number of jobs per worker (100$\times\#$ CPUs) executed on a varying number of engines. The tests were performed on five different compute configurations: the "MIST" OpenStack private cloud (top row), Amazon EC2 (2nd row), HP public cloud (3rd row), a hybrid cloud deployment with the \packagename~controller in the HP cloud and workers in Amazon EC2 cloud (4th row), and a hybrid cloud deployment with the \packagename~controller in the Amazon EC2 cloud and workers in the HP cloud (5th row).  We executed each test with the \emph{SharedStorage}, \emph{PersistentStorage} and \emph{No-Storage} methods of \emph{molnsutil}.  For the MIST cloud we also executed benchmarks of Hadoop MapReduce of the same workflow for comparison.
}
\label{fig:benchmarks}
\end{figure}

\section{Discussion}
\label{sec:Discussion}

The issue of reproducibility in scientific computation is both important and difficult to address. \packagename~constructs a templated software environment including virtualization of the computational infrastructure and the IPython notebooks contain all the code necessary to construct the models and execute the analysis workflows; thus we believe that our system holds promise to allow for easy reproduction and verification of simulation results. In particular, there is no need for a modeler to manage multiple formats of the models, or to develop code or input files specific to a particular HPC environment as all of the information is contained within the notebooks. This reduces the burden on the practitioner to learn specific computing systems, and removes the error prone and technically involved process of scaling up a computational experiment in a way that allows for collaborative analysis of the simulation results. All in all, we believe that \packagename~showcases a scientific software design that has the potential to greatly increase productivity for computational scientists.        

The current version of \packagename~makes the spatial stochastic simulation package PyURDME available as a service. PyURDME was designed from the ground up as a cloud-ready package, but in such a way that it does not rely on \packagename~for its use.  Naturally, a modeling process may want to rely on other simulation packages as well. \packagename's automatic and templated configuration of the environment can easily be extended to make other tools available in the IPython notebooks, provided that they can be accessed from the IPython environment (which is not restricted to Python code). We believe that PyURDME showcases a good design to follow for other simulation packages to benefit from this cloud delivery model. It is our hope that the \packagename~platform will grow to include a larger ecosystem of spatial and non-spatial simulation software to facilitate for practitioners to compare tools and to choose the best one for the task at hand.

We have chosen to focus our efforts in facilitating model development on constructing a programmatic interface; hence use of the service requires basic capabilities in Python programming knowledge.  The principal target user group is computational biologists that have basic knowledge of programming in a scripting language. 
By specifying models as compact Python programs, \packagename~and PyURDME join a community of computational software packages, such as PySB \cite{pysb},
whose objective is to utilize high-level, descriptive programmatic concepts to create intuitive, extensible, and reusable models that integrate advanced software engineering tools and methods to distribute and manage the computational experimental process.

From a computer science perspective, the traditional tradeoffs between interactivity and large-scale computational experiments that motivated the development of \packagename~are not unique to this particular application. Looking at scientific computing in general, applications often follow a traditional black-box execution model in which the results of the computation can be procured after the complete execution process. Such workflows have proven to be 
successful both for simple and complex applications. Queuing based job schedulers such as Torque/PBS which are typical on university clusters have been the driving 
force behind this approach. However, lack of interactivity is one of the empirical drawbacks of the
black-box execution approach. 
The cloud paradigm changes the way resources are offered, and  
therefore it is vital to change the traditional black-box execution model of scientific applications 
to support more interactivity, something that will enhance productivity, prevent wastage of computational 
resources and allow inducing knowledge on-the-fly to further optimize the ongoing analysis process. 
The issues of traditional computational workflows have been addressed within  specialized application domains. Galaxy \cite{giardine2005galaxy} provides an interactive platform that combines the existing genome wide annotations database with online analysis tools 
that enables running complex scientific queries and visualization of results. 
A commercial service, PiCloud \footnote{PiCloud is now at http://www.multyvac.com}\cite{picloud}, provided a service for distributing computation on cloud computing resources. The Control Project \cite{hellerstein1999interactive} at Berkeley focuses on a general purpose 
interactive computing techniques to enhance the human computer interaction for massive 
dataset analysis and thus provides an effective control over information. This project offers 
online aggregation, emulation and visualization and rapid data-mining tools. 
\cite{ragan2013collaborative} presents a similar approach based on StarCluster\cite{MITStarCluster} 
and IPython notebooks for multi-task computing model for reproducible biological insights. \packagename~brings the new style of IT model that the above projects represent to the domain of quantitative modeling in systems biology.  

Finally, StochSS (www.stochss.org) is a cloud-computing application developed by the present authors that aims at integrating solvers for many different model levels ranging from ordinary differential equations to spatial stochastic simulations. In contrast to \packagename, the present StochSS application emphasizes ease-of-use and targets biology practitioners with limited or no programming experience. This is reflected by a graphical web user interface (WebUI) to support modeling and a very high abstraction level for interacting with compute resources. In future work, the \packagename~platform will be consumed as-a-service within the StochSS application as an alternative to the UI-assisted modeling approach, when the user becomes more and more comfortable with quantitative modeling.

In conclusion, we present \packagename: a cloud computing virtual appliance for computational biology experiments.  It has the capability to create computational clusters from a heterogeneous set of public and private cloud computing infrastructure providers, and is bundled with the \emph{molnsutil} package to organize distributed parallel computational workflows. It uses an IPython notebook user interface designed to enable interactive, collaborative and reproducible scientific computing.
We also present PyURDME, a software package for modeling and simulation of spatial stochastic systems.  It features an intuitive and powerful model description API based on Python objects, efficient handling of complex geometries with FEniCS/Dolfin \cite{LoggMardalEtAl2012a}, fast stochastic solvers, and an extensible framework for development of advanced algorithms \cite{urdme, Drawert2010}.
Additionally, we demonstrated the capabilities of \packagename~with a computational biology study of yeast polarization.
Finally, we demonstrate shareability and reproducibility by including all the IPython notebooks used in the writing of this manuscript 
as supplemental information, and we also distribute them as examples in the \packagename~software.

\section*{Acknowledgements}
\label{sec:Acknowledgements}
We would like to acknowledge Benjamin B. Bales for useful discussions on the design and Stefan Hellander for helpful comments on the manuscript.

\section*{Author Contributions}
BD and AH developed the software;
BD and AH conceived and designed the study; 
BD, MT, ST, and AH performed the experiments and data analysis; 
all authors wrote the manuscript.

\section*{Conflict of interest}

The authors declare that they have no conflict of interest.

\bibliographystyle{siam} %
\bibliography{molns}

\end{document}